\documentstyle[12pt,epsfig]{article}
\begin{document}
\begin{center}
{\bf NUCLEON TENSOR CHARGES IN THE  CHIRAL QUARK--SOLITON MODEL}
\footnote{Talk given at 7th International Conference on the Structure of
Baryons, Santa Fe, New Mexico, 3-7 Oct 1995.} \\
\vspace{5mm}
HYUN-CHUL KIM, MAXIM V. POLYAKOV
\footnote{On leave of absence from Petersburg Nuclear Physics
Institute, Gatchina, St. Petersburg 188350, Russia}
and KLAUS GOEKE \\
\vspace{5mm}
{\it Institute for  Theoretical  Physics  II, \\
P.O. Box 102148, Ruhr-University Bochum, \\
D--44780 Bochum, Germany} \\
\vspace{5mm}
\end{center}
\date{December 1995}

\begin{abstract}
We investigate the singlet $g_T^{(0)}$  and isovector $g_T^{(3)}$
tensor charges of the nucleon, which are deeply related to
the first moment of the leading twist transversity
quark distribution $h_1(x)$, in the
chiral quark-soliton model.  With rotational $O(1/N_c) $
corrections taken into account, we obtain $g_T^{(0)}=0.69$ and
$g_T^{(3)}=1.45$ at a low normalization poin of several hundreds MeV.
 Within the same approximation and
parameters the model yields
$g_A^{(0)}=0.36$, $g_A^{(3)}=1.21$ for axial charges
and correct octet--decuplet mass
splitting.  We show how the chiral quark-soliton model interpolates
between the nonrelativistic quark model and the Skyrme model.
\end{abstract}

The nucleon tensor charges are defined as a nucleon forward matrix element
\cite{JaffeJi91}
\begin{equation}
\langle N| \bar{\psi}_f  \sigma_{\mu\nu} \psi_f |N \rangle = g_T^{f}
\bar{U}\sigma_{\mu\nu} U,
\label{tensor_charge_definition}
\end{equation}
where $U(p)$ is a standard Dirac spinor and
$\sigma_{\mu\nu}=\frac i2[\gamma_\mu, \gamma_\nu]$ in notation of Bjorken
and Drell~\cite{BD}.
The tensor charges, as shown by
Jaffe and Ji \cite{JaffeJi91}, are related to the first moment of
the transversity quark distribution $h_1(x)$:
\begin{equation}
\int_0^1dx (h_1(x)-\bar{h}_1(x))=g_T^{f},
\label{first_moment}
\end{equation}
where $f$ is a flavor index ($f=u,d,s,\cdots$).

Our aim is to calculate the tensor charges (\ref{tensor_charge_definition})
in the chiral quark--soliton model
($\chi$QSM, often called the semibosonized
Nambu---Jona-Lasinio model) at a low normalization point of several
hundreds MeV.

The $\chi$QSM has been successful in reproducing
the static properties of the baryons such as
the octet-decuplet mass splitting
\cite{DiPePo,BloDiaGoePetPobPark},
axial charges\cite{BochumSPbOsakaTokio,BloPolGoe,BloPra}
and electro.m. form factors~\cite{ChrPobNuclPhys,Kim2}.
The baryon in this model is regarded as a bound state of
$N_c$  quarks  bound by a non-trivial chiral field configuration.
Such a semiclassical picture of baryons can be justified in the
$N_c\rightarrow \infty$ limit in line with more general
arguments by Witten~\cite{Witten}.
 A remarkable virtue of $\chi$QSM
is that the model interpolates between the nonrelativistic quark
model(NRQM) and the Skyrme model~\cite{Michal}.
In particular, due to such an interplay,
it enables us to examine
the dynamical difference between the axial and tensor charges
of the nucleon.

In order to calculate the tensor charges given by
eq.(\ref{tensor_charge_definition})
we employ the effective QCD partition function folowing
from the instanton picture of QCD in the limit of low momenta.  It
is given by a functional integral over
pseudoscalar and quark fields~\cite{DyPe1}:
\begin{equation}
{\cal Z} = \int {\cal D}\Psi {\cal D}\bar\Psi
{\cal  D}\pi^A\:
\exp \left( i\int d^4x \bar\Psi iD\Psi \right),
\label{Partfunc}
\end{equation}
where $iD$ denotes the Dirac differential operator
\begin{equation}
iD\;=\;(- i \rlap{/}{\partial} + Me^{i\pi^A\tau^A\gamma_5}),
\label{Diracop}
\end{equation}
and
$M$ is the dynamical quark mass which arises as a result of the
spontaneous chiral symmetry breaking and is momentum-dependent. The
momentum dependence of $M$ introduces the natural ultra-violet cut-off
(inverse average instanton size $1/\rho \sim 600\ \mbox{MeV}$)
\cite{DyPe1} for the theory given by eq. (\ref{Partfunc})
and simultaneously brings a renormalization
scale to the model.

  One can relate the hadronic matrix element
eq. (\ref{tensor_charge_definition}) to a correlation
function:

\begin{equation}
\langle 0 | J_{B}(\vec{x},T)  \bar{\psi} \sigma_{\mu\nu} \tau^a
\psi J^{\dagger}_{B}(\vec{y},0)|0 \rangle
\label{corf}
\end{equation}
at large Euclidean time $T$ with baryon current $J_B$  constructed
from quark fields and having nucleon quantum numbers.
The correlation function (\ref{corf}) can be calculated in the effective
chiral quark model defined by eq.(\ref{Partfunc}) using $1/N_c$ expansion.
The related technique can be found in \cite{DiPePo,ChrPobNuclPhys}.
Here we give a result for the tensor charges
to the next to leading order of the $1/N_c$ expansion:
\begin{equation}
g_T^{(0)}= \frac{\alpha}{I}, \quad
g_T^{(3)}=  \beta + \frac{\delta}{I},
\label{delta}
\end{equation}
where $\alpha$, $\beta$, $\delta$ and $I \sim N_c$ can be found
in ref.\cite{KimPolGoeke}.

The $\chi$QSM interpolates between
NRQM and the Skyrme one, {\em i.e.} in the limit of small soliton size
it reproduces the results of NRQM, whereas in the opposite limit of
large soliton size it mimics the Skyrme model.
In the limit of large soliton size (large constituet quark mass),
one can easily find \cite{KimPolGoeke} that
$\alpha \sim (M R_0)^{2} $, $I \sim (MR_0)^3$ and
$\beta,\delta \sim M R_0$.  Therefore, the ratio of the tensor charges
$g_T^{(0)}/g_T^{(3)}\sim 1/(M R_0)^2$
is sizably reduced in the limit of large soliton size, while the
analogous analysis of the axial charges \cite{BloPolGoe,Michal} gives
even much stronger suppression in the ratio
$g_A^{(0)}/g_A^{(3)}\sim 1/(MR_0)^6$.  This observation
of the different behaviors between the axial and tensor charges
leads to a conclusion that the tensor charges might deviate from
axial ones remarkably.
In the limit of $M R_0 \to 0$, $\chi$QSM
corresponds to NRQM and yields:
$g_T^{(0)}=g_A^{(0)}=1$,
$g_T^{(3)}=g_A^{(3)}=(N_c+2)/3 $ (derivation for axial charges see
ref.\cite{Michal})
\footnote{Note that it is of great importance
to take into account the rotational $1/N_c$ corrections
( $\delta$ contribution in eq.~(\ref{delta})) to
derive this result in $O(N_c^0)$ order.}.

  In  figure~1 we show the dependence of the tensor and axial
charges on the soliton size.
We see that axial and tensor
charges starting from the same values of $(N_c+2)/3\approx 1.67$ for the
isovector case and $1$ for the singlet one at small soliton size have
qualitatively different behavior for larger $MR_0$ --- the dependence of
the tensor charges on soliton size is weaker than corresponding dependence
of the axial charges. This qualitative difference is in accordance with
the asymptotics of the charges in large soliton size considered above.

\vspace{0.8cm}
\centerline{\epsfysize=2.7in\epsffile{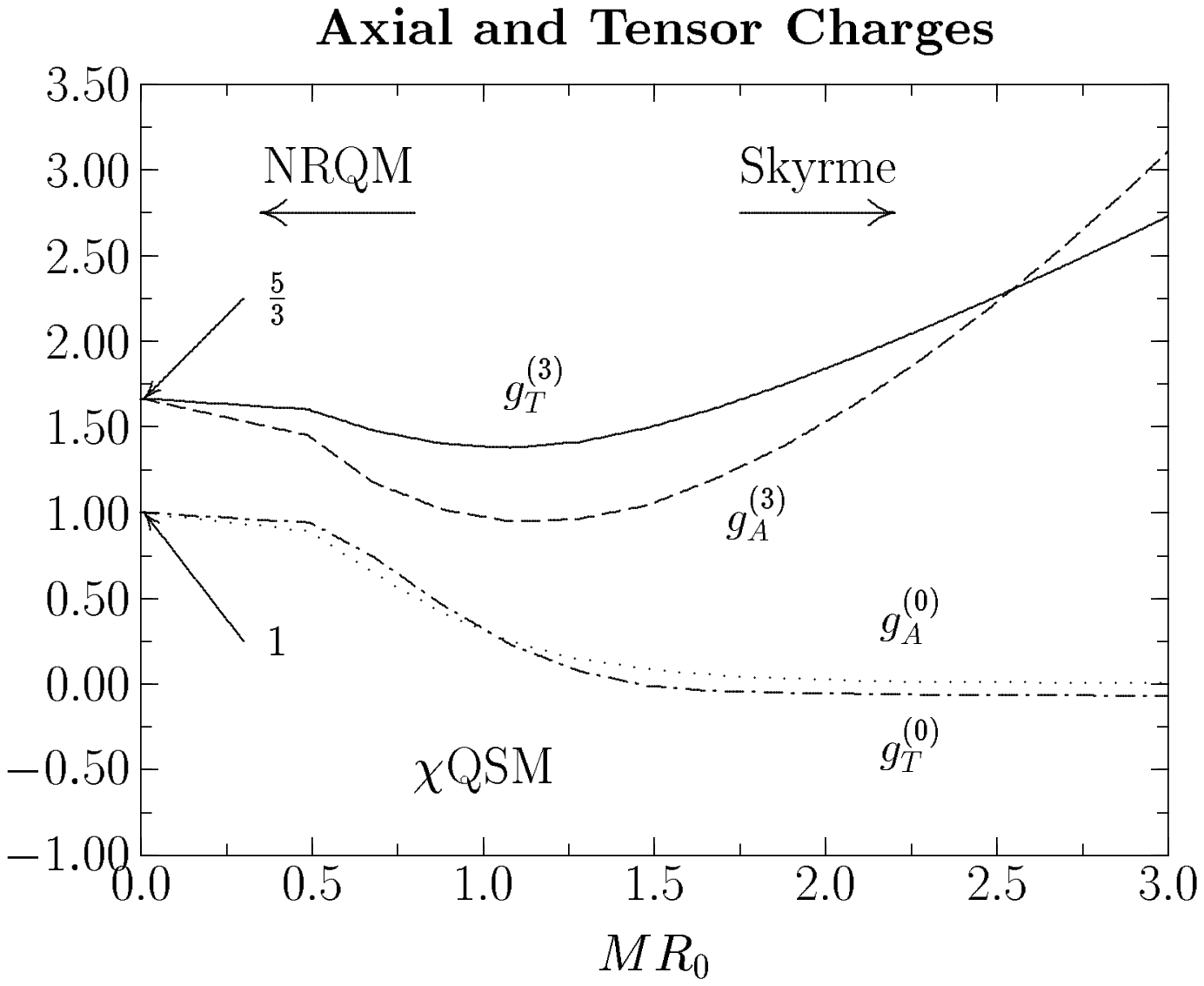}}\vskip4pt
\begin{center}
\parbox{14cm}{\footnotesize {\bf Fig. 1}:
The dependence of the axial and tensor charges on the
soliton size.
The solid curve represents the $g^{(3)}_{T}$, while the
dashed curve draws the $g^{(3)}_{A}$.  The dot-dashed curve denotes the
$g^{(0)}_{T}$, whereas the dotted curve is for the $g^{(0)}_{A}$.
The small arrows stand for the values of $g^{(3)}_{T}=g^{(3)}_{A}=5/3$ and
$g^{(0)}_{T}=g^{(0)}_{A}=1$ in NRQM, respectively.  The large
arrows denote NRQM and Skyrme limit of the present model.}
\end{center}
We have calculated the tensor charges
for  $M=420$~MeV. At this mass the model
reproduces with good accuracy many nucleon observables -- octet-decuplet
mass splitting \cite{BloDiaGoePetPobPark}, isospin splittings in baryon
 octet and decuplet \cite{BloPraGoePhysRev}, singlet axial charge
\cite{BloPolGoe,BloPra}, magnetic moments, isovector axial charge
\cite{BochumSPbOsakaTokio} and
electromagnetic form factors \cite{ChrPobNuclPhys,Kim2}.
Using accurate Kahana--Ripka method \cite{KaRi} for diagonalization
of the Dirac operator, we got:
\begin{equation}
g_T^{(3)}\approx 1.45, \qquad  g_T^{(0)}\approx 0.69.
\label{ten_num}
\end{equation}
We find that the obtained results are close to those in the bag model
\cite{JaffeJi91} and consistent with QCD sum rule calculations
of refs.~\cite{Ji_prep,IofKho}.
Using the same technique and parameters of the model one obtains the
following values of the axial charges:
\begin{equation}
g_A^{(3)}\approx 1.21, \qquad  g_A^{(0)}\approx 0.36.
\label{ax_num}
\end{equation}

It is worth noting that a dependence of the tensor charges on the
normalization point is rather weak:

\begin{equation}
g_T^{(f)}(\mu)=
\biggl(\frac{\alpha_s(\mu)}{\alpha_s(\mu_0)}\biggr)^{\frac{4}{27}}
g_T^{(f)}(\mu_0),
\label{evol}
 \end{equation}
as $\mu \to \infty $ the tensor charges slowly vanish. One can
use this equation to recalculate the tensor charges at higher
normalization points using the values of tensor
charges (\ref{ten_num}) at low normalization point.
 Due to the weak dependence of the tensor charges on the
normalization point, we do not need to know precisely the value
of the initial normalization point.

We would like to thank Chr. Christov and T. Watabe for helpful
discussions and comments.
This work has partly been supported by the BMFT, the DFG
and the COSY--Project (J\" ulich).  The work of M.P. is supported
in part by grant INTAS-93-0283.

\end{document}